\documentclass[11pt,a4paper]{article}

\usepackage[margin=2.2cm]{geometry}
\usepackage{amsmath,amssymb,bm}
\usepackage{newtxtext,newtxmath}
\usepackage{microtype}
\usepackage{booktabs,tabularx,array}
\usepackage{graphicx}
\usepackage{caption}
\usepackage{xcolor}
\usepackage{cite}
\usepackage{siunitx}
\usepackage[hidelinks]{hyperref}
\usepackage{setspace}

\newcommand{\vect}[1]{\bm{#1}}
\newcommand{\Pvec}{\vect{P}}
\newcommand{\grad}{\bm{\nabla}}
\newcommand{\dd}{\mathrm{d}}
\newcommand{\ii}{\mathrm{i}}
\newcommand{\eps}{\varepsilon}
\newcommand{\epsz}{\varepsilon^{0}}
\newcommand{\rev}[1]{{\color{black}#1}}

\graphicspath{{./}}

\title{\textbf{Mechanistic Phase-Field Modelling of Woven-Domain Formation in Ferroelectric Material}}
\author{P G Kubendran Amos\thanks{Corresponding author: \href{mailto:prince@nitt.edu}{prince@nitt.edu}}\\
\small Theoretical Metallurgy Group\\
\small Department of Metallurgical and Materials Engineering,\\
\small National Institute of Technology Tiruchirappalli, Tiruchirappalli 620015, Tamil Nadu, India}
\date{}

\begin{document}
\maketitle

\begin{abstract}
A recent experimental study reports the spontaneous formation of a three-dimensional woven ferroelectric domain fabric in bulk KTN:Li during sufficiently slow cooling through the ferroelectric transition. 
The present work develops a nondimensional phase-field framework to examine a physically plausible route by which such a state can emerge. 
A large-area two-dimensional model first resolves the cooling-rate-dependent formation of a frustrated crossing precursor by combining first-order ferroelectric thermodynamics, electrostatic and elastic interactions, compositional modulation, and effective charge screening.
Slow cooling produces a persistent population of charge-associated crossings, whereas the corresponding fast-cooling pathway does not. 
A three-dimensional extension then incorporates cubic gradient anisotropy, flexoelectric coupling, and strain-gradient regularization to examine whether the planar precursor can develop a genuine woven topology. Within the explored nondimensional parameter regime, slow cooling produces a directly resolved geometrical separation and exchange of depth ordering between the intersecting wall families, whereas the corresponding fast-cooling pathway does not. This distinction is reproduced across independent realisations and remains robust to timestep refinement and changes in computational domain size.
The simulations suggest that slow cooling provides a kinetic window for topological selection and that global charge relaxation can coexist with strong local charge concentration at surviving crossings. 
The model is not calibrated to material-specific KTN:Li coefficients and does not reproduce the experimentally observed low-temperature disappearance of the woven state.
It is therefore intended as a mechanistic, rather than quantitatively predictive, description. 
The resulting trends provide experimentally testable signatures for identifying the processes that accompany woven-domain formation.
\end{abstract}


\section{Introduction}

Ferroelectric materials are crystalline solids that develop a spontaneous electric polarization below a characteristic transition temperature and possess two or more polarization states that can be reoriented by an applied electric field. 
This switchable order parameter is strongly coupled to strain and, depending on composition and crystal symmetry, gives rise to large dielectric, piezoelectric, pyroelectric, and nonlinear optical responses.
These coupled functionalities have made ferroelectrics important to materials engineering for non-volatile memories, sensors, actuators, transducers, capacitors, and electro-optic devices \cite{Dawber2005,Wang2019}.
Equally important, a ferroelectric crystal is generally not described by a single uniform polarization state, but by a domain microstructure whose morphology is determined by competition between the local thermodynamic driving force, domain-wall energy, electrostatic interactions, elastic compatibility, defects, and external constraints.
Domain walls and more complex polar textures can themselves possess properties that are absent from the surrounding domains, making the control of domain topology an increasingly important route to functional-material design \cite{Nataf2020,Hong2026}.

Against this background, the recent observation by Xin \textit{et al.} of a spontaneously formed woven domain fabric in bulk KTN:Li represents an unusual extension of ferroelectric domain organization \cite{Xin2026}. 
During slow cooling below approximately $0.4~\mathrm{K\,min^{-1}}$, a pre-existing domain lattice reorganizes near $T_C-2~\mathrm{K}$ into an interlaced three-dimensional network whose lattice spacing is approximately twice the compositional striation period. 
Individual crossings exhibit different local over/under arrangements, the resulting configuration depends on thermal history, and charged domain-wall segments are embedded within the network.
The woven state persists over a finite temperature interval and can be locally manipulated optically, suggesting a route towards distributed, topologically robust information storage rather than storage in isolated domains or individual topological defects \cite{Xin2026,Hong2026}. 
From a materials perspective, however, the observation also poses several mechanistic questions. 
Why does weaving appear only when the crystal is cooled sufficiently slowly? How can highly charged wall intersections survive while the surrounding polarization field continues to relax?
\rev{What converts an initially planar crossing network into a genuinely three-dimensional over/under arrangement, and can that transformation be identified from a direct geometrical separation of the intersecting wall families rather than from projected morphology alone?} 
The experimentally observed relation between the woven-domain spacing and the growth striation further indicates that intrinsic domain-wall physics is being selected within a pre-existing mesoscale modulation. 
Resolving these questions is important both for understanding the phenomenon and for identifying thermal, electrical, or microstructural parameters through which the woven state might eventually be controlled.

Phase-field modelling is well suited to this problem because it follows the spatially continuous polarization field while allowing domain walls and their intersections to evolve without prescribing their final geometry. 
In the established ferroelectric phase-field framework, Landau thermodynamics is coupled to gradient, electrostatic, and elastic energies and the polarization evolves according to time-dependent Ginzburg--Landau kinetics \cite{Chen2008,Wang2019}. 
Early three-dimensional simulations by Hu and Chen demonstrate that elastic and dipolar interactions are central to the formation and arrangement of $90^{\circ}$ and $180^{\circ}$ domains \cite{Hu1998}, while Wang \textit{et al.} show how coupled electrical and mechanical interactions govern ferroelectric/ferroelastic switching pathways \cite{Wang2004}. 
Subsequent work extends this framework to non-trivial wall and topological structures. Gu \textit{et al.} demonstrate through phase-field modelling that flexoelectric coupling generated by stress gradients can alter the internal polarization structure of ferroelectric domain walls \cite{Gu2014}. 
Hong \textit{et al.} show that polar-vortex lattices can be stabilized by a balance among electrostatic, elastic, and polarization-gradient energies and that this balance selects a finite geometric length scale \cite{Hong2017}. 
Together, these studies establish that phase-field modelling can connect mesoscopic domain morphology to the competition among specific physical interactions, which is precisely the level at which the woven-domain problem is posed. 
The present problem differs, however, in that the experimentally relevant object is not an isolated wall texture or a conventional vortex lattice, but a history-dependent network of charged crossings whose defining topology requires an explicitly three-dimensional resolution.

\rev{Here, a two-stage phase-field approach is used to examine this mechanism. 
A large-area two-dimensional simulation first identifies the cooling-rate-dependent, charge-associated crossing precursor, after which a three-dimensional formulation tests whether the same pre-weave lattice develops a directly resolved through-thickness wall separation under slow cooling while remaining geometrically unresolved under fast cooling.} 
The coefficients are treated in nondimensional form and the objective is therefore to identify a physically consistent mechanism and experimentally testable trends, rather than to provide a parameter-free quantitative prediction for KTN:Li.

\section{Phase-field model and numerical implementation}
\label{sec:model}

The modelling strategy is designed to separate two questions that are experimentally intertwined but computationally distinct. 
First, can the observed thermal history select a large population of frustrated and charge-stabilized domain crossings? 
Second, once such a crossing network exists, can an initially planar domain lattice develop a genuine three-dimensional over/under arrangement and retain that topology? 
The first question is addressed using a large $256\times256$ two-dimensional simulation, which provides sufficient area for cooling-rate statistics but cannot represent one wall passing above another.
A $50\times50\times50$ three-dimensional phase-field model, in which all three polarization components are retained, is used to address the second.
The three-dimensional simulation is therefore \textit{a mechanistic model} for weaving, whereas the two-dimensional simulation is used as a \textit{large-area precursor} and cooling-rate test.
Accordingly, all coefficients reported below are dimensionless effective parameters. 
Moreover, the simulations are consequently intended to test a physically plausible mechanism and its robustness.

\subsection{Three-dimensional phase-field formulation}
\label{subsec:3dmodel}

The local ferroelectric state is represented by the polarization vector $\Pvec=(P_x,P_y,P_z)$. 
Retaining $P_z$ is essential even though the dominant experimental variants are largely in plane, since a true woven crossing requires two wall families that coincide in projection to occupy different depths. 
The total free energy functional is written as
\begin{equation}
\mathcal{F}=\int_V\left(f_{\mathrm L}+f_{\mathrm{grad}}+f_{\mathrm{el}}+f_{\mathrm{flex}}+f_{\mathrm{sg}}+f_{\mathrm{es}}\right)\,\dd V,
\label{eq:Ftotal_man}
\end{equation}
where the terms respectively describe local ferroelectric thermodynamics, domain-wall energy, electrostrictive elastic compatibility, flexoelectric coupling, strain-gradient regularization, and electrostatics. 
This decomposition, through each contribution, represents a distinct physical penalty or driving force acting on the wall network.

The local ferroelectric thermodynamics are described using a sixth-order Landau potential,
\begin{align}
f_{\mathrm L}={}&\frac{\alpha(\vect r,t)}{2}P^2-\frac{B}{4}P^4+\frac{C}{6}P^6 \\
&+A_{\mathrm{tet}}\left(P_x^2P_y^2+P_y^2P_z^2+P_z^2P_x^2\right),
\label{eq:landau_man}
\end{align}
with $P^2=P_x^2+P_y^2+P_z^2$, $B=C=1$, and $A_{\mathrm{tet}}=0.30$.
The negative quartic and positive sixth-order terms give a first-order transition. This formulation permits paraelectric and ferroelectric states to coexist metastably over a finite interval, thereby preserving thermal-history dependence close to the transition. 
The positive tetragonal anisotropy penalises simultaneous occupation of several Cartesian components and favours the six variants $\pm P_x$, $\pm P_y$, and $\pm P_z$. 
For a single-component state, the finite-polarization branch follows
\begin{equation}
P_s^2=\frac{B+\sqrt{B^2-4C\alpha}}{2C}.
\label{eq:Ps_man}
\end{equation}
For the present representative choice of $B=C=1$, the finite-polarization extremum disappears at $\alpha=0.25$ while the paraelectric and ferroelectric minima have equal free energy at $\alpha=0.1875$. 
The woven-domain formation simulations are initiated at $\alpha_0=0.15$, within this bistable range, rather than in a regime where the system is forced through a single continuous minimum.

The experimentally observed compositional striation is represented as a spatial modulation of the quadratic Landau coefficient,
\begin{equation}
\alpha(\vect r,t)=\alpha_0(t)-\Delta\alpha\cos\left(\frac{2\pi x}{\lambda_s}\right),
\label{eq:alpha_mod_man}
\end{equation}
with $\Delta\alpha=0.04$ and $\lambda_s=5$ grid cells. 
Accordingly, neighbouring compositional bands reach equivalent local thermodynamic conditions at slightly different values of the global thermal variable $\alpha_0$. 
The imposed five-cell striation is one-half of the ten-cell pre-weave superperiod used below. 
In the present non-dimensional formulation, $\alpha_0$ is a thermal control parameter and is not converted directly into $\alpha(T)$.

Domain walls are assigned a cubic, rather than isotropic, gradient energy so that longitudinal, transverse, and mixed polarization gradients can carry different energetic costs. 
In Fourier space, the corresponding contribution to the functional derivative is written as
\begin{equation}
\widehat{\left(\frac{\delta F_{\mathrm{grad}}}{\delta P_i}\right)}
=
K_{ij}(\vect{k})\widehat{P_j},
\label{eq:grad_man}
\end{equation}
where
\begin{align}
K_{ii}&=G_{11}k_i^2+G_{44}\sum_{j\ne i}k_j^2,\\
K_{ij}&=(G_{12}+G_{44})k_i k_j,\qquad i\ne j.
\end{align}
The gradient coefficients are specified in nondimensional form as
$G_{11}=0.70$, $G_{12}=-0.03$, and $G_{44}=0.02$. 
These values were arrived at following an initial isotropic formulation, in which a single gradient penalty suppressed distinctions between competing three-dimensional wall-deformation modes. 
Accordingly, $G_{11}=0.70$ retains the characteristic gradient-energy scale of the isotropic formulation and therefore preserves approximately the same overall domain-wall stiffness. 
The additional coefficients $G_{12}$ and $G_{44}$ were subsequently introduced through a restricted parameter exploration around this baseline. Their values were chosen as a weak anisotropic correction within the explored parameter range that maintained numerically stable domain walls while allowing the long-wavelength three-dimensional instability, arising from flexoelectric coupling and regularized by the strain-gradient contribution, to develop without an accompanying unphysical short-wavelength instability. 
In particular, the smaller value $G_{44}=0.02$ reduces the energetic cost of transverse wall deformation, while $G_{12}=-0.03$ introduces a weak mixed-gradient coupling between differently oriented polarization gradients and thereby modifies their cooperative deformation. 
Once this stable regime was identified, these coefficients were held fixed for all subsequent cooling, independent-realization, and persistence simulations. 
Physically, the gradient coefficients, together with the local Landau potential, influence the domain-wall width and stiffness, while their anisotropy controls the energetic cost of changing wall orientation and the coupling between spatial gradients of different polarization components.

Electrostatic effects enter as the curved, head-to-head, or tail-to-tail wall segments carry bound charge $\rho_b=-\grad\!\cdot\!\Pvec$. 
An effective screening-charge field $\rho_f$ is therefore introduced and the electrostatic potential obtained from the periodic Poisson equation
\begin{equation}
-\nabla^2\phi=\rho_b+\rho_f.
\label{eq:poisson_man}
\end{equation}
The electrostatic contribution is weighted by $q_{\mathrm{es}}=0.30$. 
The screening charge is not assumed to compensate bound charge instantaneously; instead, it evolves according to the phenomenological screening-relaxation equation
\begin{equation}
\frac{\partial\rho_f}{\partial t}=-\gamma_c(\rho_f+\rho_b)+D_c\nabla^2\rho_f,
\label{eq:charge_man}
\end{equation}
with $\gamma_c=0.10$ and $D_c=0.02$. 
While the first term drives local compensation, the second smooths strong spatial variations. 
This finite screening time ensures a wall can move, curve, or become trapped while the charge distribution is still evolving.

Polarization is coupled to mechanical deformation through electrostriction.
The spontaneous strain is written
\begin{align}
\epsz_{xx}&=q_Q\left[Q_{11}P_x^2+Q_{12}(P_y^2+P_z^2)\right],\\
\epsz_{yy}&=q_Q\left[Q_{11}P_y^2+Q_{12}(P_x^2+P_z^2)\right],\\
\epsz_{zz}&=q_Q\left[Q_{11}P_z^2+Q_{12}(P_x^2+P_y^2)\right],\\
\epsz_{xy}&=q_Q Q_{44}P_xP_y,\quad
\epsz_{xz}=q_Q Q_{44}P_xP_z,\quad
\epsz_{yz}=q_Q Q_{44}P_yP_z,
\label{eq:eigenstrain_man}
\end{align}
using $Q_{11}=1.0$, $Q_{12}=-0.35$, $Q_{44}=0.75$, and an overall electrostrictive scale $q_Q=0.30$. 
The elastic energy reads
\begin{equation}
f_{\mathrm{el}}=\frac{1}{2}C_{ijkl}(\eps_{ij}-\epsz_{ij})(\eps_{kl}-\epsz_{kl}),
\label{eq:elastic_man}
\end{equation}
with homogeneous cubic stiffnesses $C_{11}=2.0$, $C_{12}=0.8$, and $C_{44}=0.6$. 
These values define a mechanically stable cubic medium and establish the relative energetic importance of elastic compatibility. 
However, it is vital to note that, given the mechanistic focus of the present work, they are not intended to represent elastic moduli in GPa.

Though electrostriction alone produced coherent domains, it did not provide a sufficiently strong route for the planar crossing lattice to become unstable through the thickness. 
A flexoelectric coupling is therefore included between strain and the symmetric polarization-gradient tensor $S_{ij}=(\partial_jP_i+\partial_iP_j)/2$,
\begin{equation}
f_{\mathrm{flex}}=-f_{\mathrm{flexo}}\eps_{ij}S_{ij},
\label{eq:flex_man}
\end{equation}
with $f_{\mathrm{flexo}}=\textcolor{black}{14.0}$. 
Since a strong flexoelectric coupling can otherwise favour an unphysical short-wavelength instability, it is accompanied by a strain-gradient penalty
\begin{equation}
f_{\mathrm{sg}}=\frac{\ell_{\mathrm{sg}}^2}{2}C_{ijkl}\,\partial_m\eps_{ij}\,\partial_m\eps_{kl},
\label{eq:sg_man}
\end{equation}
with $\ell_{\mathrm{sg}}=\textcolor{black}{3.0}$ grid cells. 
This length suppresses very short mechanical wavelengths while leaving longer wavelength distortions relatively unaffected.
In Fourier space, mechanical equilibrium $\partial_j\sigma_{ij}=0$ is solved at every polarization step as
\begin{equation}
\widehat{u_i}(\vect k)=\frac{\Gamma^{-1}_{ij}(\vect k)b_j(\vect k)}{1+\ell_{\mathrm{sg}}^2k^2},
\label{eq:mech_man}
\end{equation}
where $\Gamma_{ij}=C_{ikjl}k_k k_l$ is the cubic acoustic tensor and $b_i=-\ii k_jT_{ij}$ contains both the electrostrictive eigenstress and the flexoelectric polarization-gradient source. 
The factor $(1+\ell_{\mathrm{sg}}^2k^2)^{-1}$ is therefore the mechanism that regularizes short-wavelength elastic response without removing the long-range compatibility interaction.

The polarization field is advanced by time-dependent Ginzburg--Landau dynamics,
\begin{equation}
\frac{\partial P_i}{\partial t}=-L_P\frac{\delta\mathcal{F}}{\delta P_i}+\xi_i(\vect r,t),
\label{eq:tdgl_man}
\end{equation}
with $L_P=1$. 
A weak zero-mean Gaussian perturbation with amplitude $8\times10^{-4}$ is used only during the near-onset formation stage to break exact translational and reflection symmetries.
It is removed during subsequent cooling and holding. 
The noise therefore does not prescribe a woven geometry, but permits otherwise symmetry-equivalent crossings to select different histories.

The principal parameters of the three-dimensional simulation are summarized in Table~\ref{tab:3dparams}. 
The values should be read as one internally consistent nondimensional parameter set that places the system close to the experimentally relevant instability.
In particular, the flexoelectric coefficient and strain-gradient length are effective model parameters. The two parameters are varied within a restricted range while all other free-energy coefficients are kept fixed, allowing the regime associated with the experimentally observed slow-cooling instability to be identified.

\begin{table}[ht]
\centering
\caption{Nondimensional parameters used in the three-dimensional simulation.}
\label{tab:3dparams}
\small
\begin{tabularx}{\textwidth}{>{\raggedright\arraybackslash}p{2.6cm} >{\centering\arraybackslash}p{1.7cm} X}
\toprule
Parameter & Value & Role in the model \\
\midrule
$B$, $C$ & 1.0, 1.0 & First-order sixth-order Landau thermodynamics. \\
$A_{\mathrm{tet}}$ & 0.30 & Tetragonal polarization anisotropy. \\
$G_{11}$, $G_{12}$, $G_{44}$ & 0.70, $-0.03$, 0.02 & Cubic domain-wall gradient coefficients. \\
$C_{11}$, $C_{12}$, $C_{44}$ & 2.0, 0.8, 0.6 & Homogeneous cubic elastic stiffness ratios. \\
$Q_{11}$, $Q_{12}$, $Q_{44}$ & 1.0, $-0.35$, 0.75 & Electrostrictive strain ratios. \\
$q_Q$ & 0.30 & Overall electrostrictive coupling strength. \\
$q_{\mathrm{es}}$ & 0.30 & Electrostatic interaction weight. \\
$\gamma_c$, $D_c$ & 0.10, 0.02 & Charge-compensation rate and charge diffusivity. \\
$f_{\mathrm{flexo}}$ & \textcolor{black}{14.0} & \textcolor{black}{Flexoelectric strain--polarization-gradient coupling in the validated geometric-crossover branch.} \\
$\ell_{\mathrm{sg}}$ & \textcolor{black}{3.0} & \textcolor{black}{Strain-gradient regularization length in the validated geometric-crossover branch (grid cells).} \\
$\Delta\alpha$, $\lambda_s$ & 0.04, 5 & Amplitude and wavelength of the compositional/$T_C$ modulation. \\
\bottomrule
\end{tabularx}
\end{table}

\subsection{Three-dimensional thermal pathway and identification of weaving}
\label{subsec:3dprotocol}

The three-dimensional simulation deliberately starts from the experimentally motivated pre-weave domain lattice rather than attempting to reproduce every stage of domain nucleation from the paraelectric state. 
This choice isolates the central mechanistic question: whether a pre-existing planar crossing lattice can become unstable to depth ordering. 
Two diagonal wall families are constructed as
\begin{align}
f_1(x,y)&=\tanh\left[\frac{\sin\{2\pi(x+y)/\Lambda_d\}}{s_w}\right],\\
f_2(x,y)&=\tanh\left[\frac{\sin\{2\pi(x-y)/\Lambda_d\}}{s_w}\right],
\end{align}
with a domain superperiod $\Lambda_d=10$ cells and wall-sharpness parameter $s_w=0.65$. 
The initial polarization then reads
\begin{equation}
P_x=\frac{P_s}{2}(f_1+f_2),\qquad
P_y=\frac{P_s}{2}(f_1-f_2),\qquad
P_z=0,
\label{eq:init_man}
\end{equation}
where $P_s$ follows Eq.~\eqref{eq:Ps_man} at $\alpha_0=0.15$, giving $P_s\simeq0.903$. 
The lattice is relaxed briefly while enforcing uniformity through $z$ so that numerical discretization transients do not themselves create a three-dimensional mode.

{\color{black}
The three-dimensional simulation uses the same free-energy functional, but the instability is initialized and assessed geometrically. 
A restricted exploration of only the flexoelectric coupling and strain-gradient regularisation identifies the branch $f_{\mathrm{flexo}}=14.0$ and $\ell_{\mathrm{sg}}=3.0$, while all other free-energy and kinetic coefficients are retained. 
The pre-relaxed planar polarisation texture is then given a small wall-following displacement,
\begin{equation}
(u_x,u_y)=u_0\,(T_x,T_y)
\sin\left(\frac{2\pi m z}{N_z}+\varphi\right),
\qquad u_0=0.03,\quad m=5,
\label{eq:geompert_man}
\end{equation}
where $(T_x,T_y)$ is the normalised in-plane template obtained from the gradients of the two pre-existing wall families and $\varphi$ is chosen independently for each realization. 
The displaced texture is obtained by sampling the relaxed field at $(x-u_x,y-u_y)$ for each $z$ plane. 
The amplitude $u_0=0.03$ grid cell corresponds to $0.3\%$ of the ten-cell domain superperiod and is far below the one-cell criterion subsequently used to identify a resolved cross-over. 
The perturbation therefore supplies only a candidate through-thickness wavelength and does not construct the final woven geometry or prescribe which wall family occupies a given depth.

Fast and slow simulations start from this same perturbed state and span the same interval $\alpha_0=0.15\rightarrow-0.12$. 
The slow pathway uses 45 near-onset formation steps with the weak stochastic forcing retained, followed by 40 deterministic cooling steps and a 150-step final hold. 
The fast pathway traverses the corresponding stages in 8, 5, and 30 steps, respectively. 
Thus the distinction between the two paths is the time available for the polarization, elastic, and screening fields to reorganize, rather than a change in the thermodynamic endpoints. 
The production simulations use a $50\times50\times50$ grid with unit spacing, periodic boundary conditions in all three directions, and $\Delta t=0.0035$. 
Spatial derivatives, Poisson electrostatics, and elastic equilibrium are evaluated spectrally. 
Seeds 101, 203, and 307 are used for independent realisations.

Weaving is not a cross-correlation lag. 
At each projected crossing, the zero-contours of $f_1=P_x+P_y$ and $f_2=P_x-P_y$ are located with sub-grid interpolation as a function of depth, giving wall-position curves $d_1(z)$ and $d_2(z)$. 
Their relative displacement,
\begin{equation}
\Delta d(z)=d_1(z)-d_2(z),
\label{eq:deltad_man}
\end{equation}
is the direct geometric measure of over/under separation. 
A crossing is counted as a resolved woven cross-over only when $\Delta d(z)$ changes sign through the thickness and exceeds both $+1$ and $-1$ grid cell. 
The same fixed criterion is applied to every projected crossing in both thermal histories.
No crossing is selected or classified by visual inspection. 
Numerical robustness is assessed by halving the timestep to $\Delta t=0.00175$ while doubling the integration counts, and by repeating the simulation on $40^3$ and $60^3$ cells while retaining the ten-cell characteristic depth wavelength.
}

\subsection{Large-area two-dimensional precursor and cooling-rate simulation}
\label{subsec:2dmodel}

The $256\times256$ model is used to establish whether cooling rate can select a statistically robust population of frustrated, charge-associated crossings over an area substantially larger than the $50^3$ three-dimensional cell. 
Since a two-dimensional field cannot represent a true over/under topology, a ``locked X'' in this simulation is interpreted primarily as a kinetically stabilized planar precursor. 
The in-plane polarization $\Pvec=(P_x,P_y)$ is described by
\begin{equation}
f_{\mathrm L}^{2D}=\frac{\alpha}{2}(P_x^2+P_y^2)+\frac{b}{4}(P_x^4+P_y^4)+\frac{c}{2}P_x^2P_y^2,
\label{eq:landau2d_man}
\end{equation}
with $b=1.0$ and $c=2.5$.
The positive mixed term favours the four axis-oriented variants $\pm P_x$ and $\pm P_y$.
Domain-wall energy is represented by an isotropic coefficient $\kappa=0.70$, while the spatial thermal/compositional modulation has amplitude $A_{T_C}=0.10$. 
To avoid a periodicity mismatch in the larger cell, the striation is made exactly commensurate with the domain using 26 periods, giving $\lambda_s^{2D}=256/26=9.846$ cells.

The reduced model also contains two quenched fields that represent weak frozen heterogeneity. 
A smooth random vector bias $\vect h=(h_x,h_y)$ enters through $-D_h(h_xP_x+h_yP_y)$ with $D_h=0.11$ and Gaussian correlation width 1.2 cells. 
A scalar field $\eta(\vect r)$ introduces weak local variant anisotropy through $A_\eta\eta(P_y^2-P_x^2)$ with $A_\eta=0.04$ and correlation length 2 cells. 
These fields remain fixed throughout a run. 
The same nondimensional elastic constants, electrostrictive ratios, electrostatic weight, and charge-screening coefficients used in the precursor branch were retained: $C_{11}=2.0$, $C_{12}=0.8$, $C_{44}=0.6$, $Q_{11}=1.0$, $Q_{12}=-0.35$, $Q_{44}=0.75$, $q_Q=q_{\mathrm{es}}=0.30$, $\gamma_c=0.10$, and $D_c=0.02$.

Charge accumulation is allowed to reduce local domain-wall mobility through
\begin{equation}
L_{\mathrm{eff}}(\vect r,t)=\frac{1}{1+\beta\left(|\rho_f|/\rho_0\right)^2},
\label{eq:mob2d_man}
\end{equation}
with $\beta=2.2$ and $\rho_0=0.10$, so that the two-dimensional TDGL equation becomes $\partial_tP_i=-L_{\mathrm{eff}}\,\delta\mathcal F_{2D}/\delta P_i$.
This term is a reduced kinetic representation of charge-assisted pinning: once screening charge accumulates at a strongly charged junction, the local polarization field becomes less mobile.
It is not used as a substitute for three-dimensional topology; the latter is assessed only in the $50^3$ simulation.

The two-dimensional grid includes $256^2$ nodes with unit spacing and timestep $\Delta t=0.018$. 
For each seed, the same initial polarization and quenched disorder are subjected to two thermal histories spanning the same control-parameter interval, $a_0=0.04$ to $-0.95$: slow cooling used 2160 steps, whereas fast cooling used 180 steps. 
The slow-cooled state is subsequently held for 1500 deterministic steps to test whether the selected crossings survived after the thermal ramp. 
Seeds 101, 203, and 307 are used for the large-domain statistics.
A separate timestep check with $\Delta t=0.009$ and twice as many integration steps is used to verify that the crossing population is not an explicit-timestep artifact. 
A planar junction is counted as a locked crossing only when it retained a strict four-arm X geometry and its local charge-dependent mobility falls below the prescribed locking criterion. 
Table~\ref{tab:2dparams} summarizes the parameters specific to this reduced large-area simulation.

\begin{table}[ht]
\centering
\caption{Parameters specific to the $256\times256$ two-dimensional precursor simulation.}
\label{tab:2dparams}
\small
\begin{tabularx}{\textwidth}{>{\raggedright\arraybackslash}p{2.8cm} >{\centering\arraybackslash}p{1.9cm} X}
\toprule
Parameter & Value & Role in the model \\
\midrule
$b$, $c$ & 1.0, 2.5 & In-plane quartic self-energy and mixed-variant penalty. \\
$\kappa$ & 0.70 & Isotropic 2D domain-wall gradient coefficient. \\
$A_{T_C}$ & 0.10 & Amplitude of periodic thermal/compositional modulation. \\
$\lambda_s^{2D}$ & 9.846 & Commensurate striation period, $256/26$, in grid cells. \\
$D_h$ & 0.11 & Strength of the smooth quenched random-bias field. \\
Random-field width & 1.2 & Gaussian correlation width in grid cells. \\
$A_\eta$, $\ell_\eta$ & 0.04, 2 & Strength and correlation length of quenched variant anisotropy. \\
$\beta$, $\rho_0$ & 2.2, 0.10 & Strength and charge scale of mobility reduction. \\
Grid, $\Delta t$ & $256^2$, 0.018 & Production spatial domain and timestep. \\
Slow / fast cooling & 2160 / 180 & Steps over the same interval $a_0=0.04\rightarrow-0.95$. \\
Final hold & 1500 & Deterministic persistence test after slow cooling. \\
\bottomrule
\end{tabularx}
\end{table}

Ultimately, the two simulations deliberately make different claims. 
The $256^2$ model establishes cooling-rate selectivity, charge localization, persistence, and large-area consistency across the tested realizations of a frustrated crossing precursor, whereas the $50^3$ model tests the additional three-dimensional instability required for true over/under weaving. 
The three-dimensional simulation begins from an experimentally informed pre-weave lattice and introduces only a $0.03$-cell geometrical perturbation at the ten-cell depth wavelength. 
The resulting topology is accepted as woven only when direct zero-contour tracking demonstrates a sign-changing wall separation exceeding $\pm1$ grid cell. 
The retained $f_{\mathrm{flexo}}=14.0$, $\ell_{\mathrm{sg}}=3.0$ branch is an effective nondimensional parameter regime identified through restricted exploration.
The periodic boundaries, homogeneous elastic properties, idealized sinusoidal striation, and nondimensional coefficients further limit quantitative comparison with a real specimen. 
In addition, the experimentally observed disappearance of the woven state upon further cooling is not reproduced by the present formulation. 
The model should therefore be treated as an initial mechanistic demonstration that the coupled action of first-order ferroelectric thermodynamics, elastic compatibility, electrostatics and screening, flexoelectricity, and strain-gradient regularization can provide a physically plausible and repeatable route, within the explored parameter regime, from a pre-existing domain lattice to a history-dependent woven state.

\section{Results and discussion}
\label{sec:results_discussion}

The simulations are used to separate two stages of the experimentally observed woven-domain transition. 
The large-area two-dimensional simulation first tests whether thermal history alone can select and preserve a population of frustrated, charge-associated crossings. 
The three-dimensional simulation then asks whether such a crossing network develops a directly measurable geometric separation of the two wall families through the sample thickness, and whether that separation is selected by slow rather than fast cooling. 
This separation leaves different experimental signatures: the first stage is primarily expressed through cooling-rate dependence, junction persistence, and charge localization, whereas the second is identified by a sign-changing, finite wall-surface displacement through the thickness.

\subsection{Slow cooling selects a persistent, charge-associated precursor state}
\label{subsec:2d_results}

\begin{figure}[p]
\centering
\includegraphics[width=0.98\textwidth]{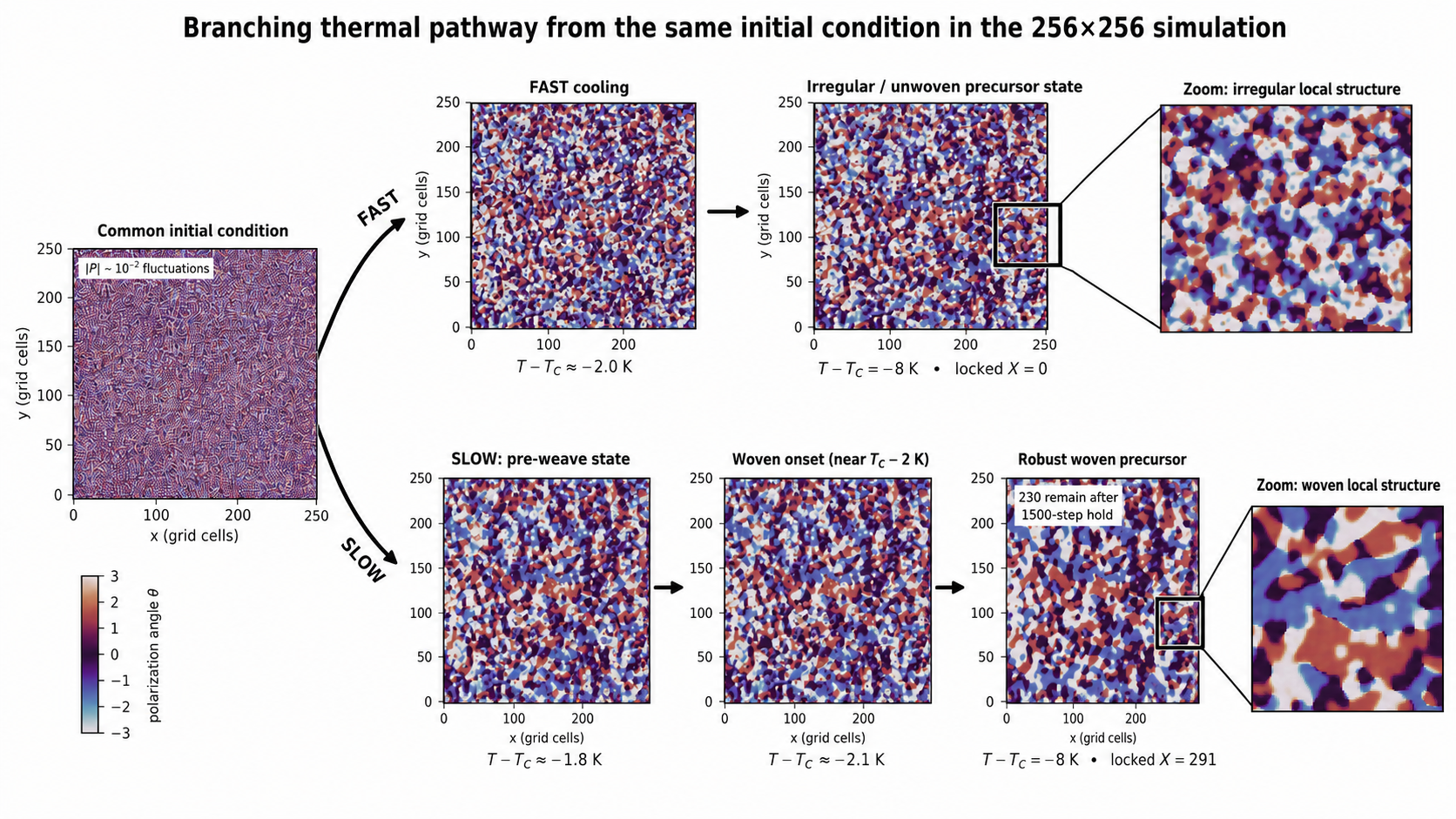}
\caption{Branching thermal pathway in the $256\times256$ precursor simulation. The same initial state is subjected to fast and slow cooling. Fast cooling produces an irregular unwoven state with no retained locked crossings, whereas slow cooling develops a pre-weave state, crossing onset near $T_C-2$~K, and a persistent frustrated-crossing population on further cooling and holding.}
\label{fig:2d_branch}
\end{figure}

Figure~\ref{fig:2d_branch} compares fast and slow cooling from the same initial polarization field and quenched heterogeneity. 
The two thermal pathways are therefore distinguished by the time available for the coupled polarization and screening fields to reorganize. 
Under fast cooling, the domain pattern remains highly irregular as the system passes through approximately $T-T_C=-2$~K and reaches $T-T_C=-8$~K without retaining a strict locked crossing. 
In contrast, slow cooling develops a recognizable pre-weave state near $T-T_C\simeq-1.8$~K, followed by the appearance of frustrated crossings close to $T-T_C\simeq-2.1$~K. 
For the representative seed 203, 291 locked crossings are present at the end of slow cooling at $T-T_C=-8$~K, whereas the corresponding fast-cooled state contains none. 
The same slow-cooled structure retains 230 crossings after a further 1500-step deterministic hold. 
Accordingly, the principal effect of slow cooling is not merely to produce a different instantaneous domain configuration, it places the system in a state from which a substantial crossing population survives even after the thermal ramp has ended.

\begin{figure}[p]
\centering
\includegraphics[width=0.97\textwidth]{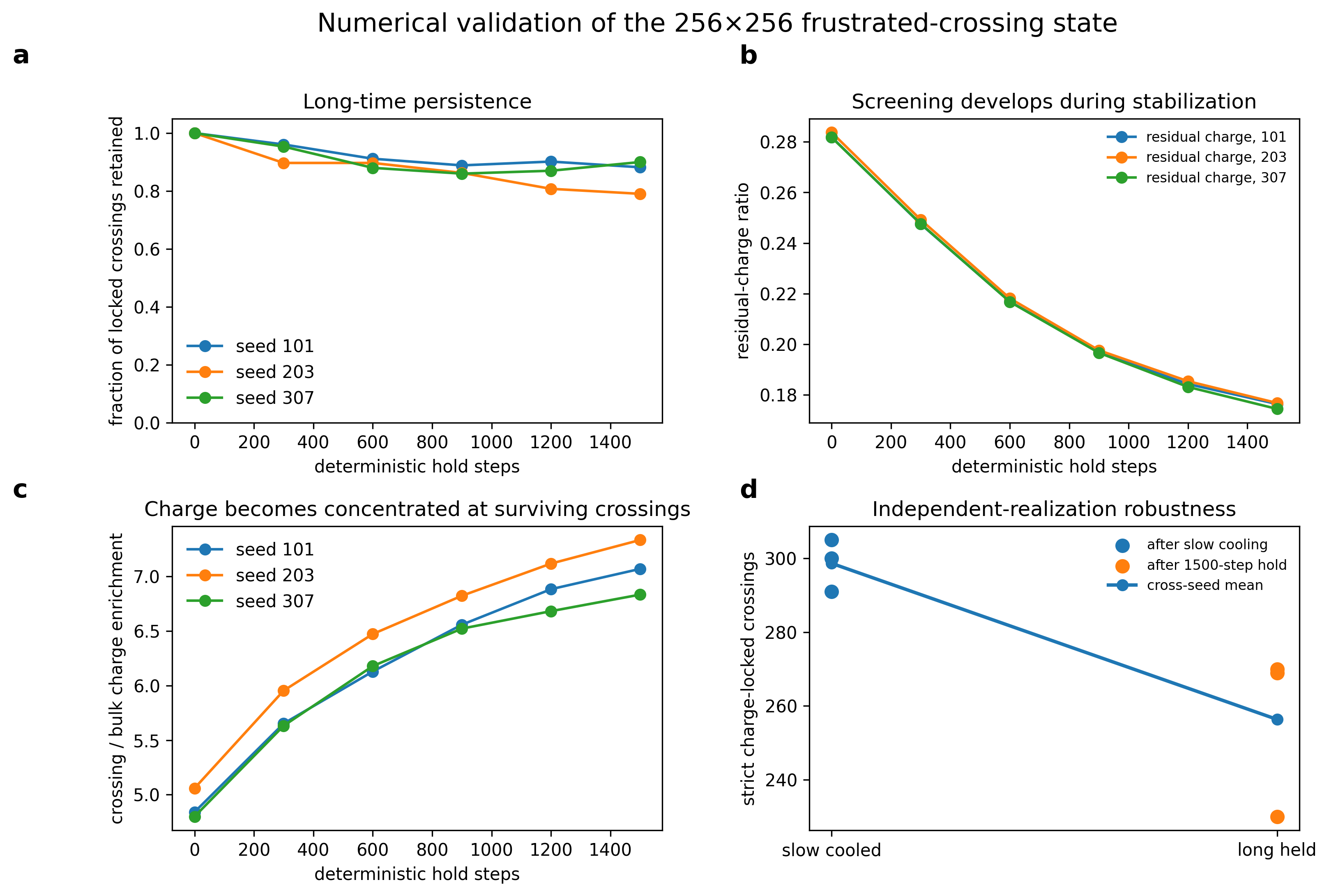}
\caption{Numerical validation of the $256\times256$ frustrated-crossing state. (a) Fraction of the initially selected crossings retained during the deterministic hold. (b) Reduction of the residual-charge ratio as screening develops. (c) Increasing charge enrichment at the crossings that survive. (d) Crossing populations obtained from three independent realizations immediately after slow cooling and after the 1500-step hold.}
\label{fig:2d_validation}
\end{figure}

The independent-realisation statistics in Fig.~\ref{fig:2d_validation} show that this behaviour is not restricted to a single random seed.
Immediately after slow cooling, seeds 101, 203, and 307 contain 305, 291, and 300 strict charge-locked crossings, respectively.
After the 1500-step hold, 269, 230, and 270 remain, corresponding to retained fractions of approximately 0.88, 0.79, and 0.90. 
The absolute number therefore changes during relaxation, but the crossing-rich state is not erased. 
The result is particularly relevant to the experimental observation of a narrow thermal window for woven-domain formation: the model indicates that the cooling rate can determine whether the evolving wall network has sufficient time to enter a long-lived crossing configuration before the surrounding domain structure coarsens or rearranges. 
Correspondingly, the experimentally observed onset near $T_C-2$~K need not be viewed simply as the temperature at which a new equilibrium morphology becomes favourable.
It can instead mark a kinetic window in which wall motion, electrostatic screening, and local pinning operate on comparable time scales and allow the crossing network to become trapped.

A more revealing feature is the simultaneous evolution of the total residual charge and the charge carried by the surviving crossings. 
During the deterministic hold, the residual-charge ratio decreases from approximately $0.28$ immediately after slow cooling to about $0.17$--$0.18$ for all three realizations [Fig.~\ref{fig:2d_validation}(b)]. 
At the same time, the ratio of charge at surviving crossings to that in the surrounding bulk increases from approximately $4.8$--$5.1$ to $6.8$--$7.3$ [Fig.~\ref{fig:2d_validation}(c)]. 
These two trends show that stabilization does not require the entire microstructure to remain strongly unscreened. 
Instead, screening progressively reduces the distributed electrostatic burden while the crossings that survive become increasingly distinguished as local charge-rich objects. 
The calculated precursor is therefore best described as a \emph{screened but locally charge-concentrated} state rather than a globally charged wall network.

This distinction suggests an experimentally accessible way of separating domain formation from electrostatic stabilization.
If charge-sensitive contrast can be measured during or after slow cooling, the model predicts that the strongest electrostatic signature should become progressively concentrated at a subset of persistent junctions even while the average unscreened charge of the domain pattern decreases. 
Consequently, an isothermal hold after slow cooling should not simply freeze the microstructure unchanged: some crossings are expected to disappear, while those that remain should become more strongly associated with localized screening charge. 
A time-resolved comparison of domain images with surface-potential, local-conductivity, or other charge-sensitive measurements would therefore test a central prediction of the reduced model more directly than morphology alone.

The two-dimensional simulation nevertheless stops at a frustrated planar crossing. 
A strict X-shaped junction can be long lived in two dimensions, but one wall cannot pass above or below the other.
The simulation therefore identifies a kinetic precursor to weaving rather than the woven topology itself. 
This distinction motivates the three-dimensional simulation, where the same competition between wall energy, electrostatics, elasticity, and screening is supplemented by the through-thickness degree of freedom required for depth ordering.

\subsection{Instability resolves planar crossings into a woven topology}
\label{subsec:3d_results}

{\color{black}
Figure~\ref{fig:3d_branch} shows fields rendered directly from $P_x$, $P_y$, and $P_z$ arrays for the validated $f_{\mathrm{flexo}}=14.0$, $\ell_{\mathrm{sg}}=3.0$ branch. 
The same geometrically perturbed pre-weave lattice is used for both thermal histories. 
The fast pathway remains close to the columnar through-thickness arrangement of the initial state, whereas the slow pathway progressively develops a stronger depth-dependent reorganization during formation, cooling, and the final hold. 
The projected $xy$ maps change less dramatically than the corresponding through-thickness state, emphasising that projected morphology alone is not a sufficient test of weaving.

\begin{figure}[p]
\centering
\includegraphics[width=0.98\textwidth]{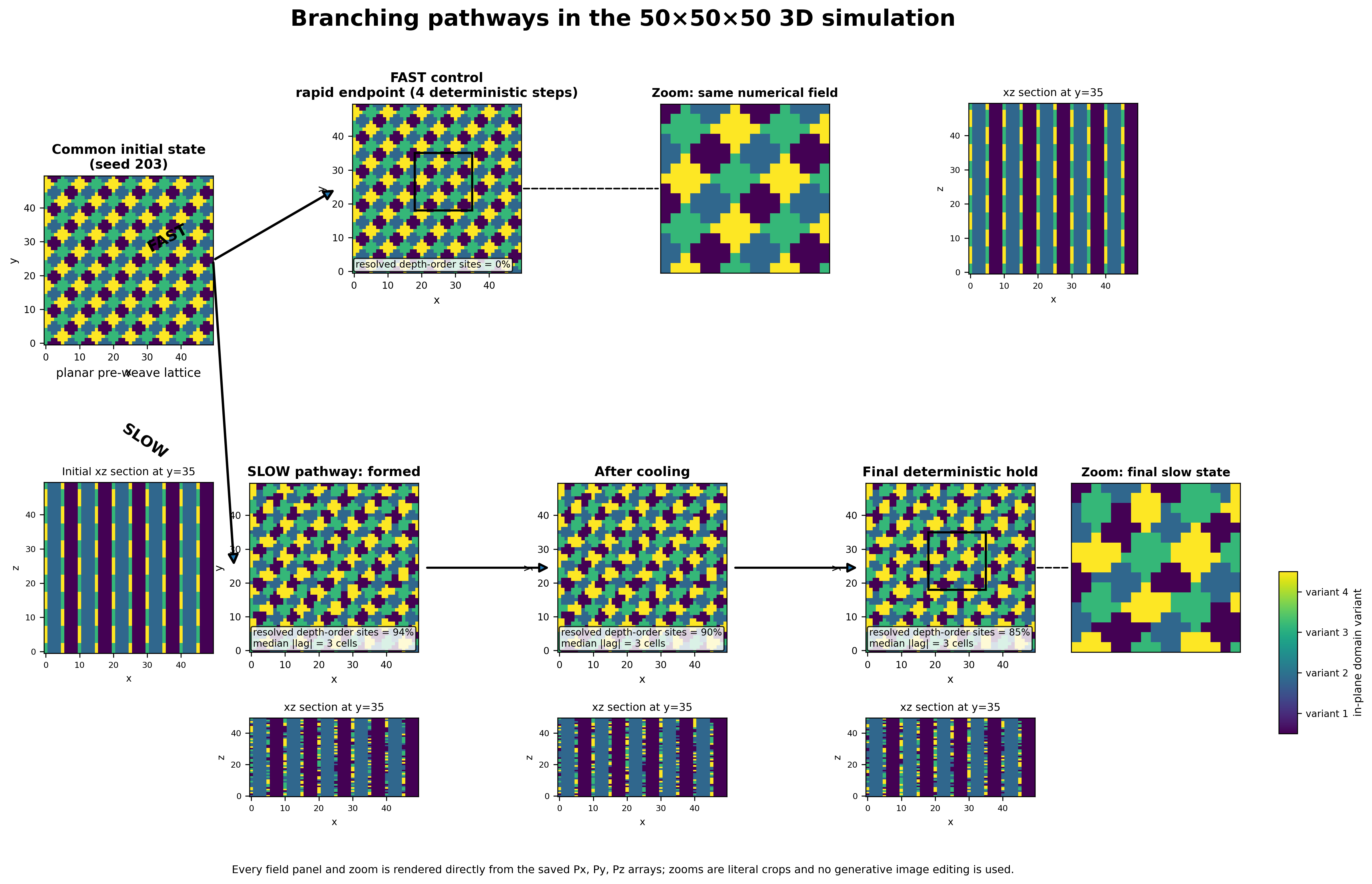}
\caption{\textcolor{black}{Branching pathways in the validated $50\times50\times50$ three-dimensional simulation. Every field panel and local crop is rendered directly from the saved polarization arrays for seed 203. The common initial state is subjected to the fast and slow thermal protocols described in the text. The fast final state remains close to the initially columnar through-thickness arrangement, whereas the slow pathway develops progressively stronger depth-dependent reorganization from formation through cooling and holding. The direct geometric criterion used to establish over/under cross-over is shown separately in Fig.~\ref{fig:3d_geometry}.}}
\label{fig:3d_branch}
\end{figure}

The geometrical distinction is quantified in Fig.~\ref{fig:3d_geometry} by tracking the two wall zero-contours. 
The representative crossing is chosen objectively as the slow-cooled crossing whose root-mean-square separation is closest to the median value over all crossings.
The identical projected crossing is then evaluated in the fast branch. 
For the fast pathway, $\Delta d(z)$ remains between approximately $-0.168$ and $+0.138$ grid cell. 
For the slow pathway at the same crossing, $\Delta d(z)$ varies from approximately $-1.114$ to $+1.129$ grid cells, demonstrating both a resolved separation and an exchange in relative depth ordering. 
This is a direct geometrical cross-over and does not rely on a correlation-based depth proxy.

\begin{figure}[p]
\centering
\includegraphics[width=0.97\textwidth]{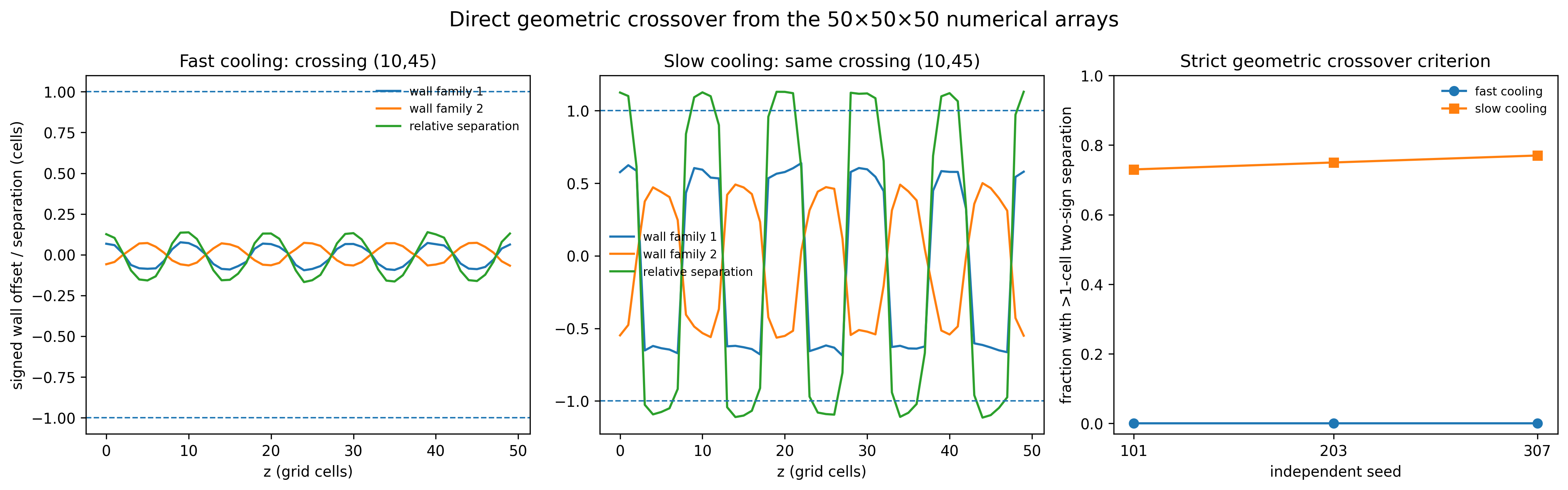}
\caption{Direct geometrical demonstration of the slow--fast crossover in the validated three-dimensional branch. The same projected crossing is analysed in the fast and slow states. The wall-location curves $d_1(z)$ and $d_2(z)$ and their relative separation $\Delta d(z)$ show that the fast state remains below the one-cell resolution criterion, whereas the slow state exhibits a sign-changing separation exceeding both $+1$ and $-1$ grid cell. Applying the same criterion to every projected crossing gives resolved-crossing fractions of $0$ for fast cooling and $0.73$, $0.75$, and $0.77$ for slow cooling in seeds 101, 203, and 307, respectively.}
\label{fig:3d_geometry}
\end{figure}

The population-level result is therefore substantially stronger than a visually selected example. 
Across the three independent realisations, none of the fast-cooled crossings satisfies the strict geometric criterion, whereas $73$--$77\%$ of the slow-cooled crossings do. 
Slow cooling consequently acts as a kinetic gate for the growth of the through-thickness displacement.
The same initial microstructure and thermodynamic interval are used, but the additional time available during the slow path allows the wall families to develop a finite depth separation. 
This behaviour suggests that the onset of weaving should be sought not only as a change in projected domain morphology but also as the emergence of a depth-dependent wall corrugation or exchange of wall order that may require cross-sectional or depth-sensitive imaging to resolve.

The numerical checks in Fig.~\ref{fig:3d_validation} show that the slow--fast distinction is not generated by the explicit timestep or by the $50^3$ box. 
Halving the timestep from $0.0035$ to $0.00175$, with the integration counts doubled to preserve the same simulated time, leaves the fast fraction at zero and changes the slow fraction only from $0.75$ to $0.79$ for seed 203. 
The corresponding median slow-state separation changes from $1.0108$ to $1.0049$ grid cells. 
Repeating the simulation on $40^3$, $50^3$, and $60^3$ domains gives slow-cooled resolved fractions of $0.719$, $0.750$, and $0.819$, respectively, while the fast fraction remains zero in every case. 
The dominant through-thickness mode changes from $m=4$ to $m=5$ to $m=6$ as the cell size increases, maintaining a physical wavelength close to ten grid cells. 
The depth wavelength is therefore not locked to a particular box mode, although its coincidence with the imposed ten-cell domain superperiod means that the present simulation does not independently predict the experimentally observed length-scale relation.

\begin{figure}[p]
\centering
\includegraphics[width=0.97\textwidth]{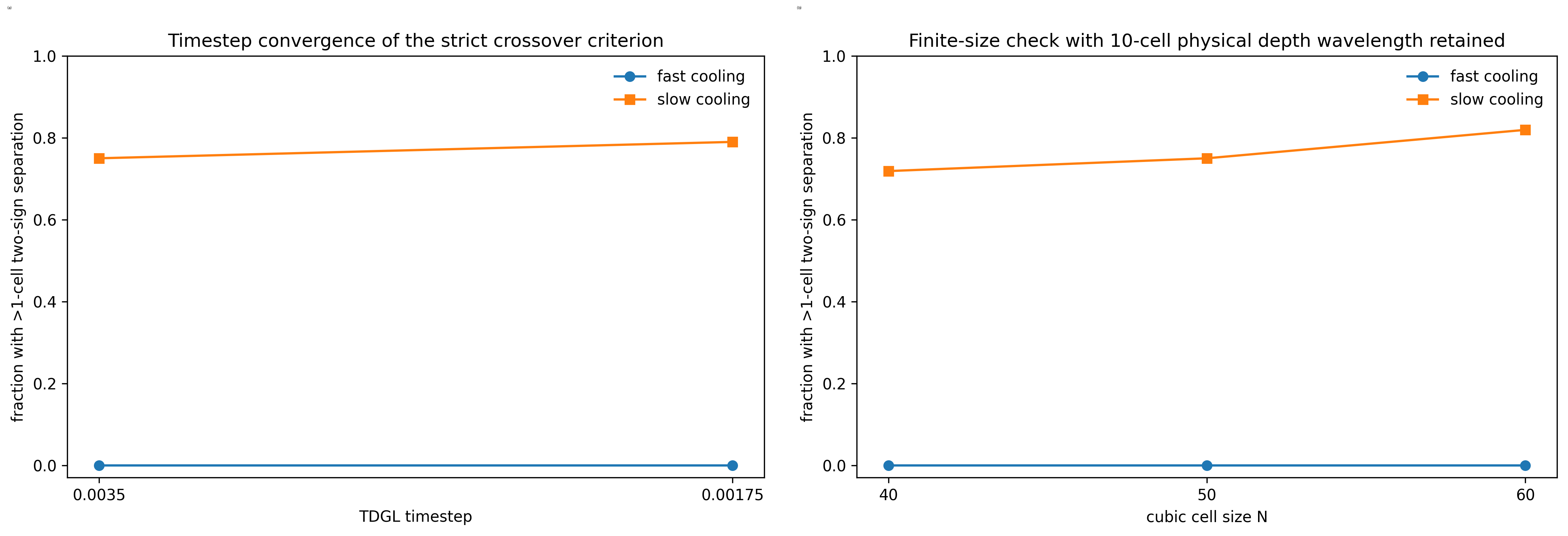}
\caption{\textcolor{black}{Numerical robustness of the direct geometric crossover. (a) Timestep refinement leaves the fast pathway at zero resolved crossings and changes the slow-cooling fraction only modestly. (b) Finite-size checks on $40^3$, $50^3$, and $60^3$ cells preserve the qualitative slow--fast distinction. The corresponding dominant modes $m=4$, $5$, and $6$ maintain an approximately ten-cell through-thickness wavelength.}}
\label{fig:3d_validation}
\end{figure}

These results refine the physical interpretation of the experimental observation. 
The two-dimensional simulations identify the slow-cooling window in which a persistent crossing precursor can be retained, while the three-dimensional simulation shows that, within the explored nondimensional parameter regime, the same thermal selectivity can control whether those crossings acquire a real over/under geometry. 
The result also explains why a woven transition may be underestimated from surface or projected imaging alone: a major part of the structural change can occur as relative wall motion through the sample thickness. 
At the same time, the values $f_{\mathrm{flexo}}=14.0$ and $\ell_{\mathrm{sg}}=3.0$ remain effective parameters selected through restricted exploration rather than independently measured KTN:Li coefficients, and the model still does not reproduce the experimentally observed disappearance of the woven state upon further cooling. 
The present simulations therefore support a mechanistic route for onset and geometrical selection of the weave, but not yet a material-specific quantitative prediction or a complete description of the low-temperature dissolution.
}

\section{Conclusion}

The present study establishes a mechanistic phase-field framework for understanding how a woven ferroelectric domain state can emerge from a pre-existing domain lattice under an appropriate thermal pathway. 
The simulations show that slow cooling does more than alter the instantaneous domain configuration: it creates a kinetic window in which frustrated crossings form, persist, and become progressively associated with localized bound charge even as the overall residual charge in the system decreases through screening.
The two-dimensional simulations therefore identify a charge-associated precursor state whose formation is strongly cooling-rate dependent, while the three-dimensional simulations demonstrate the corresponding over/under geometry directly by tracking the two wall surfaces through the thickness. Across three independent realizations, $73$--$77\%$ of the slow-cooled crossings develop a sign-changing relative wall separation exceeding $\pm1$ grid cell, whereas none of the fast-cooled crossings does so. This distinction remains under timestep refinement and finite-size checks, providing direct numerical evidence that slow cooling can act as a kinetic gate for geometrical weaving within the explored parameter regime.
These results provide several experimentally accessible insights: the relevant transition should be sought within a restricted temperature--time window during slow cooling, persistent crossings may remain locally more strongly charged than their surroundings even after substantial global screening, and the decisive structural evolution may occur as relative wall motion through the specimen thickness without producing an equally dramatic change in the projected domain pattern.
Together, these observations offer a physically plausible interpretation of the experimentally observed cooling-rate sensitivity, sudden appearance, local over/under character, and persistence of the woven-domain state.

Despite the insights offered, the present formulation remains a mechanistic rather than quantitatively predictive model. 
The material coefficients are used in nondimensional form and are not independently calibrated to the experimental system, while the computational domains, periodic boundary conditions, prescribed compositional striation, and experimentally informed initial domain lattice simplify the actual microstructure.
The retained three-dimensional branch uses effective values $f_{\mathrm{flexo}}=14.0$ and $\ell_{\mathrm{sg}}=3.0$ identified through restricted parameter exploration, and the approximately ten-cell through-thickness wavelength is not predicted independently from measured material constants.
The simulations do not yet reproduce the experimentally observed disappearance of the woven state upon further cooling. 
In addition, the effective screening-charge kinetics do not constitute a full microscopic carrier-transport description, and the present simulations cannot establish that the proposed mechanism is unique. 
Subsequent work will therefore focus on material-specific parameterization, larger three-dimensional domains, explicit treatment of realistic defects and boundary conditions, more physical charge-transport kinetics, prediction of wavelength selection from experimentally measurable quantities, and extension of the thermal pathway to capture both the formation and eventual disappearance of the woven structure.
These developments will allow the present mechanistic framework to evolve toward a quantitatively testable description of woven-domain formation in ferroelectric materials.

\section*{Acknowledgement}
The author thanks the Anusandhan National Research Foundation (ANRF) for its financial support through the Advanced Research Grant, \\
ANRF/ARG/2025/000213/ENS.

\section*{Data Availability}
In this theoretical work no data was generated.

\end{document}